\documentclass[aps,prb,showpacs,superscriptaddress,amsmath,twocolumn]{revtex4-1}

\usepackage{amsmath}
\usepackage{amssymb}
\usepackage{graphicx}

\begin{document}

\title{Sublattice asymmetry and spin-orbit interaction 
induced out-of-plane spin polarization of photoelectrons 
}

\author{P. Rakyta}
\affiliation{Department of Physics of Complex Systems,
E{\"o}tv{\"o}s University,
H-1117 Budapest, P\'azm\'any P{\'e}ter s{\'e}t\'any 1/A, Hungary
}

\author{A. Korm\'anyos}
\thanks{e-mail: a.kormanyos@lancaster.ac.uk}
\affiliation{Department of Physics, Lancaster University, Lancaster, LA1 4YB, United Kingdom}

\author{J. Cserti}
%\thanks{e-mail: cserti@complex.elte.hu}
\affiliation{Department of Physics of Complex Systems,
E{\"o}tv{\"o}s University,
H-1117 Budapest, P\'azm\'any P{\'e}ter s{\'e}t\'any 1/A, Hungary
}

%\wideabs{

\begin{abstract}
We study theoretically the effect of spin-orbit coupling and sublattice asymmetry in graphene 
on the spin polarization of photoelectrons.  
We show that  sublattice asymmetry in graphene not only opens a gap in the band 
structure but in the case of finite spin-orbit interaction it also gives rise 
to an out-of-plane spin polarization  of electrons close to the Dirac point of the 
Brillouin zone. This can be detected by measuring the spin polarization of photoelectrons and 
therefore   spin resolved photoemission spectroscopy can reveal the presence of a band gap 
even if it is too small to be observed  directly by  angle resolved photoemission spectroscopy  
because of the finite resolution of measurements or because the sample is $p$-doped.
We present  analytical and numerical calculations on the energy and linewidth dependence
of photoelectron intensity distribution and spin polarization.

\end{abstract}


\maketitle

\section{Introduction}

There is growing evidence that in addition to its extraordinary electronic properties\cite{graphene-1},  
graphene might also be an exciting  material for spintronics, a technology that would
be based on the spin of electrons rather than on their charge. 
The impetus to study spin-related phenomena in graphene comes  from 
two directions: i) the experiments of Tombros \emph{et al} (Ref.~\onlinecite{tombros}) showed 
that it was possible to 
inject spin into mono- and few layers graphene and measure spin signals in a spin-valve 
setup, and ii) the recent observation\cite{varykhalov,giant} of  band splitting in graphene 
due to spin-orbit interaction (SOI). Although the intrinsic SOI\cite{kane} is expected  
to be weak in graphene (not exceeding\cite{huertas-1,ISO-calcs,gmitra,abdelouahed}  
$\approx 50 \mu{\rm eV}$), the breaking of the inversion symmetry by 
an external electric field or by the presence of a substrate can result in a substantial
externally induced SOI. In particular, Varykhalov \emph{et al} (Ref.~\onlinecite{varykhalov}) reported 
a spin-orbit interaction induced band splitting of $\approx 13\,{\rm meV}$ in a  
quasi-free-standing  graphene on Ni(111)/Au substrate. The spin-orbit coupling
was identified as Rashba-type SOI\cite{kane,rashba} (RSOI) and it was attributed to the high nuclear
charge of the gold atoms that were intercalated between the Ni substrate and the 
graphene layer to break the strong carbon-nickel bonds and make the  graphene layer 
quasi-free-standing. The fact that gold intercalation can decouple graphene from the nickel substrate
was also supported by density functional calculations\cite{ho-kang} and that it 
may induce sizeable Rashba-type SOI was indicated by the computations of Ref.~\onlinecite{abdelouahed}. 
Furthermore,  gold intercalation was  used to decouple graphene grown on  
Ru(0001) substrate\cite{ruthenium-gap} where 
a band-gap opening at the Dirac-point of the graphene band structure was observed as well. 
The appearance of the gap was ascribed to the breaking of the  symmetry of the two carbon 
sublattices in graphene. A gap opening in the graphene band structure 
was also found when the strong nickel-graphene bonds were passivated by potassium 
intercalation\cite{gruneis}. 
Besides metal surfaces (for a review see Ref.~\onlinecite{metal-review}), 
intensive research effort, 
both theoretical\cite{seungchul,qi,pankratov} and experimental (see e.g. 
Refs.~\onlinecite{giant,qi,gierz-gold,riedl,siegel}, earlier 
developments are reviewed in Ref.~\onlinecite{sic-review}), 
is directed towards   studying graphene on SiC substrate. 
These experiments  show therefore that 
substrates  can induce SOI and/or open a band gap in monolayer graphene. 

Angle-resolved photoemission spectroscopy (ARPES) is an important experimental technique  
that provides direct information  on  the bulk and surface electronic band structure of 
solid state materials [see e.g. Refs.~\onlinecite{arpes-hightc,arpes-metal}]. 
ARPES has also become a major tool to study graphene on various 
substrates\cite{varykhalov,giant,ruthenium-gap,gruneis,gierz-gold,siegel,zhou,bostwick-1,bostwick-2,
pletikosic,sprinkle,sic-gap,s-polar}
By also measuring the spin polarization of the photoelectrons (the so called 
spin-resolved ARPES or  SARPES technique\cite{sarpes-1}) and using 
a sophisticated  data analysis method\cite{sarpes-2} one may observe band splittings 
smaller than the intrinsic linewidth of  regular ARPES experiments, providing 
a powerful tool to measure  spin resolved electronic bandstructure.     
Indeed, SARPES measurements were used in the  experiments of Refs.~\onlinecite{varykhalov,giant} to 
investigate the spin dependent band splitting in graphene.

Our work is motivated by the fact that, as mentioned above, substrates 
can induce RSOI and/or open band gap in monolayer graphene. Therefore the interplay of the 
two effects, i.e. the RSOI and sublattice asymmetry induced band gap opening may be important in some systems. 
We note that small band-gaps (by which we mean a few tens of ${\rm meV}$) are not easily detected 
by ARPES because of the finite experimental resolution ($10-50\,{\rm meV}$) and because 
of the finite intrinsic linewidth in the measurements.
We theoretically demonstrate that broken carbon sublattice symmetry coupled with RSOI 
induces a finite out-of plane spin polarization in monolayer graphene, therefore SARPES measurements 
could detect small band gaps even if conventional ARPES can not. 
We study  the constant-energy angular maps 
and the spin-resolved momentum distribution curves (MDC)\cite{sarpes-1,sarpes-2}  of photoelectrons 
as a function of initial state energy and line-broadening for finite RSOI and sublattice 
asymmetry induced band-gaps. Our work is therefore complementary to Ref.~\onlinecite{falko} in which 
a similar study was published for zero SOI and 
also to Ref.~\onlinecite{kuemmeth} which focused on the effect of RSOI on photoelectrons but  
the sublattice asymmetry was not considered and the dependence of the MDC-s on initial state energy 
and line broadening was not discussed in details. 

The rest of the paper is organized in the following way: first, in Section~\ref{sec:graphene-bands}
we show that if both RSOI and sublattice asymmetry are present then the quasiparticles in 
monolayer graphene acquire a non-zero out-of-plane spin polarization 
in a part of the Brillouin zone. We then show in Section~\ref{sec:sarpes-theor} 
how the spin polarization of quasiparticles (both in and out-of plane) is related to 
the spin polarization of photoelectrons. Using these results in Section~\ref{sec:num-sarpes} 
we  present a numerical study on the initial-state energy and intrinsic line 
broadening dependence of fixed-energy ARPES angular maps and spin-resolved MDC's and we 
point out the signatures of sublattice asymmetry. Finally, in Section~\ref{sec:summary}
we discuss the possible experimental relevance of our work and give a short summary.
Some details of the calculations in  Section~\ref{sec:sarpes-theor}   
can be found in Appendix~\ref{sec:matrix_element}.

\section{RSOI and sublattice asymmetry in graphene monolayer}
\label{sec:graphene-bands}

\begin{figure}
\includegraphics[width=0.45\textwidth]{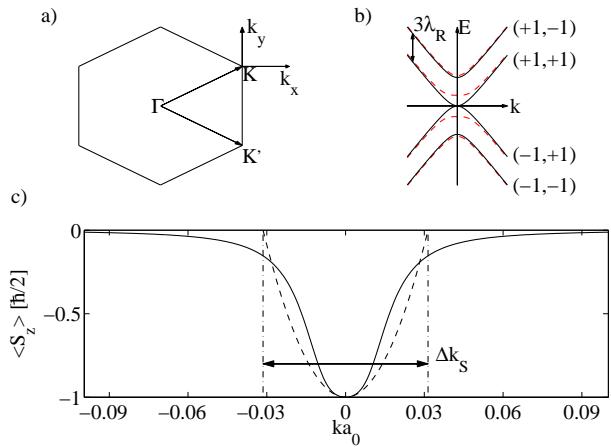}
\caption{a) Schematic of the hexagonal Brillouin zone of graphene with the 
$\mathbf{\Gamma}$ point and vectors $\mathbf{K}$, $\mathbf{K}'$ pointing to the corresponding
corners of the Brillouin zone.  The coordinate system  
that we use in the momentum  space is also shown.
b) Schematic of the energy  bands near the ${K}$ point of the BZ, 
as obtained from Eq.~(\ref{eq:spectrum}) for $\Delta=0$ (solid) and $\Delta\neq0$ (dashed). 
The energy splitting between the spin-split bands is $3\lambda_R$ for $\Delta=0$. If $\Delta\neq 0$, 
a band gap of $\Delta$ opens at $\mathbf{k}=0$. 
The ($\mu,\nu$) indeces corresponding to a given band are also indicated. 
c) solid [dashed] line: expectation value of the $z$ component of the spin as a function of 
$k=|\mathbf{k}|$ (in units of the carbon-carbon bond length $a_0$) 
in the upper valance band ($\mu=-1$, $\nu=1$)
calculated from Eq.~(\ref{eq:spin-z}) [Eq.~(\ref{eq:sz_approx})]. 
The interplay of sublattice asymmetry and RSOI leads  to a finite $\langle S_z\rangle$.
 %$z$ component of spin polarization.
The width of the peak, as defined  in the figure,  is independent of the asymmetry parameter $\Delta$ 
[see text below Eq.~(\ref{eq:sz_approx})].} 
\label{fig:Spinz}
\end{figure}

In a previous publication\cite{sajat} we showed that starting from the tight-binding Hamiltonian suggested 
in Ref~\onlinecite{kane} to describe RSOI in monolayer graphene one can arrive at the 
following Hamiltonian in the continuum limit at the ${K}$ point of the Brillouin zone (BZ):
\begin{equation}
 {H}_{RSO} = \left(\begin{smallmatrix}
                  0 & v_F \hat{p}_- & 0 & -v_{\lambda} \hat{p}_+ \\
		v_F \hat{p}_+ & 0 & 3 i\lambda_R  & 0 \\
		0 & -3 i\lambda_R  & 0 & v_F \hat{p}_- \\
                -v_{\lambda} \hat{p}_- & 0 & v_F \hat{p}_+ & 0
                 \end{smallmatrix}\right). 
\label{eq:H_RSO}
\end{equation}
[The BZ of graphene with the high 
symmetry points ${\Gamma}$, ${K}$ and ${K}'$ is shown in  Fig.~\ref{fig:Spinz}(a).] 
$H_{RSO}$ in Eq.~(\ref{eq:H_RSO}) is written in the basis 
$\{ |A\uparrow\rangle, |B\uparrow\rangle, |A\downarrow\rangle, |B\downarrow\rangle \}$ 
($\{A, B\}$ denoting the two triangular sublattice of graphene's honeycomb lattice and 
$\{ \uparrow,\downarrow \}$ is the basis in  spin Hilbert space). The parameters 
appearing in the Hamiltonian (\ref{eq:H_RSO}) are as follows: 
$v_F = 3\gamma_0 a_0/(2\hbar)$, where $a_0$ is the bond length between the carbon atoms, 
$\gamma_0$ is the hopping amplitude between next neighbour carbon
atoms,  $v_{\lambda} = 3\lambda_R a_0/(2\hbar)$, where $\lambda_R$ gives the strength of the RSOI 
in the tight-binding model of Ref~\onlinecite{kane}.  
Furthermore, $\hat{p}_{\pm} = \hat{p}_x\pm i \hat{p}_y$ and $\hat{p}_x,\hat{p}_y$ are momentum operators.
The Hamiltonian (\ref{eq:H_RSO}) differs from the Hamiltonian put forward in Ref.~\onlinecite{rashba} 
 by the terms $v_{\lambda} \hat{p}_{\pm}$. Like the terms $\pm 3i\lambda_R$ they appear because of the 
spin-orbit interaction and  they lead to trigonal warping of the bands at low energies, i.e. close to the  
${K}$ point of the BZ.  
Note that for wavenumbers  far from ${K}$ point  there is another 
kind of trigonal deformation of the bands, which  is a lattice effect [see e.g. in 
Fig.~\ref{fig:SARPES-highenergy}(b)]. 
It turns out that one can  understand\cite{sajat} all the salient features of the spin polarization at low energies 
already without taking into account the   $v_{\lambda} \hat{p}_{\pm}$ term because it 
gives higher order corrections in the wave vector $\mathbf{k}=(k_x,k_y)$ [measured from the ${K}$ point]. 
Therefore in our analytical calculations we use the following Hamiltonian:
\begin{equation}
 H_{RSO,AB} = \left(\begin{smallmatrix}
                  \frac{\Delta}{2}  & v_F \hat{p}_- & 0 & 0 \\
		v_F \hat{p}_+ & -\frac{\Delta}{2} & 3 i\lambda_R  & 0 \\
		0 & -3 i\lambda_R  & \frac{\Delta}{2} & v_F \hat{p}_- \\
                0 & 0 & v_F \hat{p}_+ & -\frac{\Delta}{2}
                               \end{smallmatrix}\right)\;. 
\label{eq:H_RSOAB}
\end{equation}
where the terms $\pm \Delta/2$  account for a possible breaking of the symmetry of 
the sublattices $A$ and $B$. %of  graphene's honeycomb lattice.

The eigenvalues of Hamiltonian (\ref{eq:H_RSOAB}) are:
\begin{equation}
\varepsilon_{\mu\nu}({\bf k}) = \frac{\mu}{2}
\sqrt{4v_F^2\hbar^2k^2 + \Delta^2 + 18\lambda_R^2 - \nu 18\mathcal{N}(k)}\;, 
\label{eq:spectrum}
\end{equation}
where  $k=|\mathbf{k}|$ and 
\begin{equation}
 \mathcal{N}(k) = |\lambda_R|\sqrt{\frac{4}{9}v_F^2\hbar^2k^2 + \lambda_R^2}\;.
\end{equation}
The index $\mu=1 (-1)$ corresponds to  conductance (valance) bands, whereas
$\nu=1$ for the low energy bands touching at $\mathbf{k}=0$ (for $\Delta=0$) and $\nu=-1$ 
for the spin split bands\cite{kane,rashba,kuemmeth} 
[a schematic of the band structure is shown in Fig.~\ref{fig:Spinz}(b)]. 
In the case of $AB$ asymmetry, i.e. for $\Delta\neq 0$ a gap   opens in the spectrum 
at the Dirac-point  ($\mathbf{k}=0$).

The RSOI  leads to a particular spin polarization of the bands\cite{kane,kuemmeth}.  
The expectation value of the three components of the quasiparticle spin 
in an eigenstate $|\Psi^{\mu,\nu}(\mathbf{k})\rangle$ of the Hamiltonian (\ref{eq:H_RSOAB})
can be calculated as:
\begin{equation}
 ^{\mu, \nu}\langle S_{x,y,z}\rangle = {\rm Tr} \left( Q^{\mu, \nu}\hat{S}_{x,y,z} \right)\;.
\label{eq:mean-bloch-spin}
\end{equation}
Here $Q^{\mu, \nu}({\bf k})=|\Psi^{\mu,\nu}(\mathbf{k})\rangle \langle\Psi^{\mu,\nu}(\mathbf{k})|$ 
is a $4\times4$ projector %onto the $(\mu,\nu)$ subspace 
and  a convenient way to  calculate these projectors   
can be found in Appendix~\ref{sec:matrix_element}.
The operator $\hat{S}_{x,y,z}$ is given by $\hat{S}_{x,y,z} = \frac{\hbar}{2}(I_2 \otimes {\sigma}_{x,y,z})$ 
where $I_2$ is the  
$2\times2$ identity matrix acting in the pseudospin space and 
$\sigma_{x,y,z}$ are Pauli matrices acting in the quasiparticles' spin space.
The expectation  values  of the spin components are found to be (in units of $\frac{\hbar}{2}$):
\begin{equation}
 ^{\mu, \nu}\langle S_x\rangle = \nu\frac{-2\hbar v_F k_y\lambda_R}{3\mathcal{N}(k)}\;,
\quad ^{\mu, \nu}\langle S_y\rangle = \nu\frac{2\hbar v_F k_x\lambda_R}{3\mathcal{N}(k)}\;,
\label{eq:spin-xy}
\end{equation}
and
\begin{equation}
 ^{\mu, \nu}\langle S_z\rangle = \mu\nu\frac{\Delta\lambda_R^2}{2\mathcal{N}(k)|\varepsilon_{\mu\nu}({\bf k})|}\;. 
\label{eq:spin-z}
\end{equation}
The  $x$ and $y$ components of the spin polarization are independent of the sublattice asymmetry 
[Eqs.~(\ref{eq:spin-xy})] and we obtain the same results as in  Refs.~\onlinecite{kane,rashba,kuemmeth,sajat} 
 i.e. the in-plane component of the spin shows rotational symmetry around 
the ${K}$ point, it is perpendicular to the wavevector ${\bf k}$ and its magnitude depends on $k$, vanishing
at $\mathbf{k}=0$. 
One can see from Eq.~(\ref{eq:spin-z}) that compared to the case of equivalent sublattices ($\Delta=0$) 
where the spin has only in-plane components for all bands\cite{kane,rashba,kuemmeth,sajat},
 the interplay of sublattice asymmetry and RSOI leads to 
finite $z$ spin polarization of electrons in the vicinity of the  ${K}$ point.
For the $\nu=1$ bands  exactly in the  ${K}$ point ($\mathbf{k}=0$) the spins are fully polarized 
and perpendicular to the graphene sheet: 
$^{\mu, 1}\langle S_z\rangle=\mu=\pm 1$, while for the $\nu=-1$ bands the spin $z$ component points 
into the opposite direction as in the $\nu=1$ bands and it can be significantly smaller: 
$^{\mu, -1}\langle S_z\rangle=-\mu\Delta/\sqrt{\Delta^2+36\lambda_R^2}$. 
(We note that $^{\mu, \nu}\langle S_z\rangle$ is the expectation value of the spin $z$ component 
\emph{averaged over a unit cell} and not on individual carbon atoms within the unit cell, 
which was discussed in Ref.~\onlinecite{ming-hao}.)   
At the ${K}'$ point, the other (inequivalent)  point of the graphene BZ where the valence and conductance
bands touch for $\Delta=0$, the
spin polarization is exactly opposite than at  the ${K}$ point, as required by the time-reversal 
symmetry.
Expanding the right hand side of Eq.~(\ref{eq:spin-z}) assuming that $\hbar v_F k \ll \lambda_R$ 
we find  for the $\nu=1$ bands:
\begin{equation}
 ^{\mu,1}\langle S_z\rangle \approx 
\mu\left(1-\frac{2 \hbar^2 v_F^2 k^2}{9\lambda_R^2}\right) 
\label{eq:sz_approx} %\\
\end{equation}
which  one can use to  give an estimate of the wavenumber range where the  spin $z$ component is non-zero. 
Fig.~\ref{fig:Spinz}(c) shows the $z$ polarization 
%for one of the valence bands, $^{-1,1}\langle S_z\rangle$, 
computed using Eq.~(\ref{eq:spin-z}) 
and  its approximation from Eq.~(\ref{eq:sz_approx}). Estimating the width of the peak
by the $k$ values where Eq.~(\ref{eq:sz_approx}) becomes zero we find 
$\Delta k_S = 3 \sqrt{2}\lambda_R/ \hbar v_F$, which is independent of the asymmetry parameter $\Delta$.
 Taking $v_F\approx 10^6{\rm m/s}$ and e.g. $\lambda_R\approx20\,{\rm meV}$  we find that 
 $\Delta k_S\approx 0.01 {\rm \AA^{-1}}$.

\section{Theoretical description of SARPES for Graphene}
\label{sec:sarpes-theor}

In this and the next section we will analyze the effects of sublattice asymmetry on the SARPES spectra  
performing both analytical and numerical calculations. 
As in most of the relevant graphene literature\cite{falko,kuemmeth,shirley}, 
we assume that the emitted photoelectrons can be characterized by a simple plane wave of momentum $\mathbf{p}$, 
spin $\sigma$ and energy $E_{\mathbf{p},\sigma} = \frac{\mathbf{p}^2}{2m_e}$ 
(however, see e.g.  Ref.~\onlinecite{s-polar} for the limitations of  this assumption). 
The flux  of photoelectrons emitted  from an initial state of momentum $\hbar(\mathbf{K}+\mathbf{k})$, 
energy  $\varepsilon_{\mu\nu}(\mathbf{k})$ in band $(\mu,\nu)$ is found to be 
  \begin{equation}
 \begin{split}
  ^{\mu,\nu}I%(\mathbf{p}_{\parallel})
  \propto &{\rm Tr}\;\bigg(\widetilde{Q}^{\mu,\nu}({\bf k})\bigg)\;\\
 &\delta_{ \mathbf{p}_{\parallel}/\hbar-(\mathbf{K}+\mathbf{k}+\mathbf{G}),\mathbf{0}}\;
 \delta(\hbar\omega + \varepsilon_{\mu\nu}(\mathbf{k})-E_{{\bf p},\sigma}-W)\;.
 \end{split}
\label{eq:intensity}
 \end{equation}
[Some details of the calculations leading to Eq.~(\ref{eq:intensity}) and  
Eq.~(\ref{eq:O_Tr}) below are given in Appendix \ref{sec:matrix_element}.]
Here $\widetilde{Q}^{\mu, \nu}({\bf k})$ is a $2\times2$ projector onto the photoelectron spinor:
\begin{equation}
 \widetilde{Q}_{ij}^{\mu, \nu}({\bf k}) = \sum\limits_{k=2i-1}^{2i}\sum\limits_{l=2j-1}^{2j} 
\left(U Q^{\mu, \nu}({\bf k})U^{\dagger} \right)_{kl}, 
\label{eq:Q-tilde}
\end{equation}
where  $Q^{\mu, \nu}$ was introduced after Eq.~(\ref{eq:mean-bloch-spin}) and the unitary matrix $U$  is given by
\begin{equation}
 U = \left(\begin{smallmatrix}
      1 & 0 & 0 & 0 \\
      0 & e^{-{\rm i}{\bf G}{\boldsymbol\tau}} & 0 & 0 \\
      0 & 0 & 1 & 0 \\
      0 & 0 & 0 & e^{-{\rm i}{\bf G}{\boldsymbol\tau}}
     \end{smallmatrix}\right)\;.
\end{equation}
Here $\mathbf{G}$ is an arbitrary reciprocal lattice vector 
 and $\boldsymbol{\tau}$ is a vector pointing from lattice site $B$ to site $A$ in the unit cell of graphene.  
In the following we will always take ${\bf G}={\bf 0}$, since  we will concentrate on one  BZ.   
The Kronecker delta in  Eq.~(\ref{eq:O_Tr}) expresses  momentum conservation ($\mathbf{p}_{\parallel}$ is 
the component of the momentum of photoelectrons parallel with the graphene surface). 
Finally, the Dirac delta function in Eq.~(\ref{eq:O_Tr}) ensures the energy conservation 
($W$ being the work function of graphene). 
We do not address the question of dynamical processes that lead to energy broadening but use 
a phenomenological approach by introducing a Lorentzian   
$\delta(\varepsilon) \rightarrow \frac{\Gamma^2}{\varepsilon^2 + \Gamma^2}$ 
(see Ref.~\onlinecite{lorentzian}) in the figures of Section~\ref{sec:num-sarpes} with the parameter
$\Gamma$ representing the value of the broadening.
To keep the formulae uncluttered we suppress henceforth the Kronecker and Dirac delta functions expressing the 
momentum and energy conservation, they should be understood to appear on the right hand side of 
Eqs.~(\ref{eq:ARPESint})-(\ref{eq:SARPES-photoe}) below.

\begin{figure}
\includegraphics[width=0.28\textwidth]{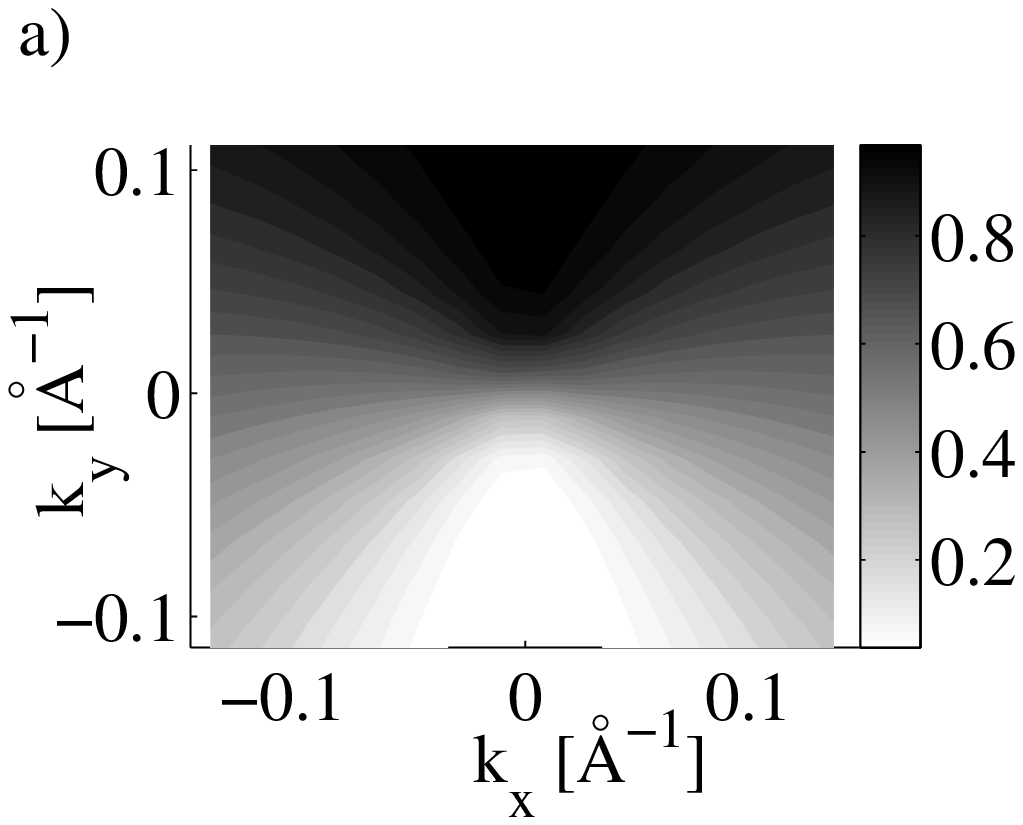}
\includegraphics[width=0.5\textwidth]{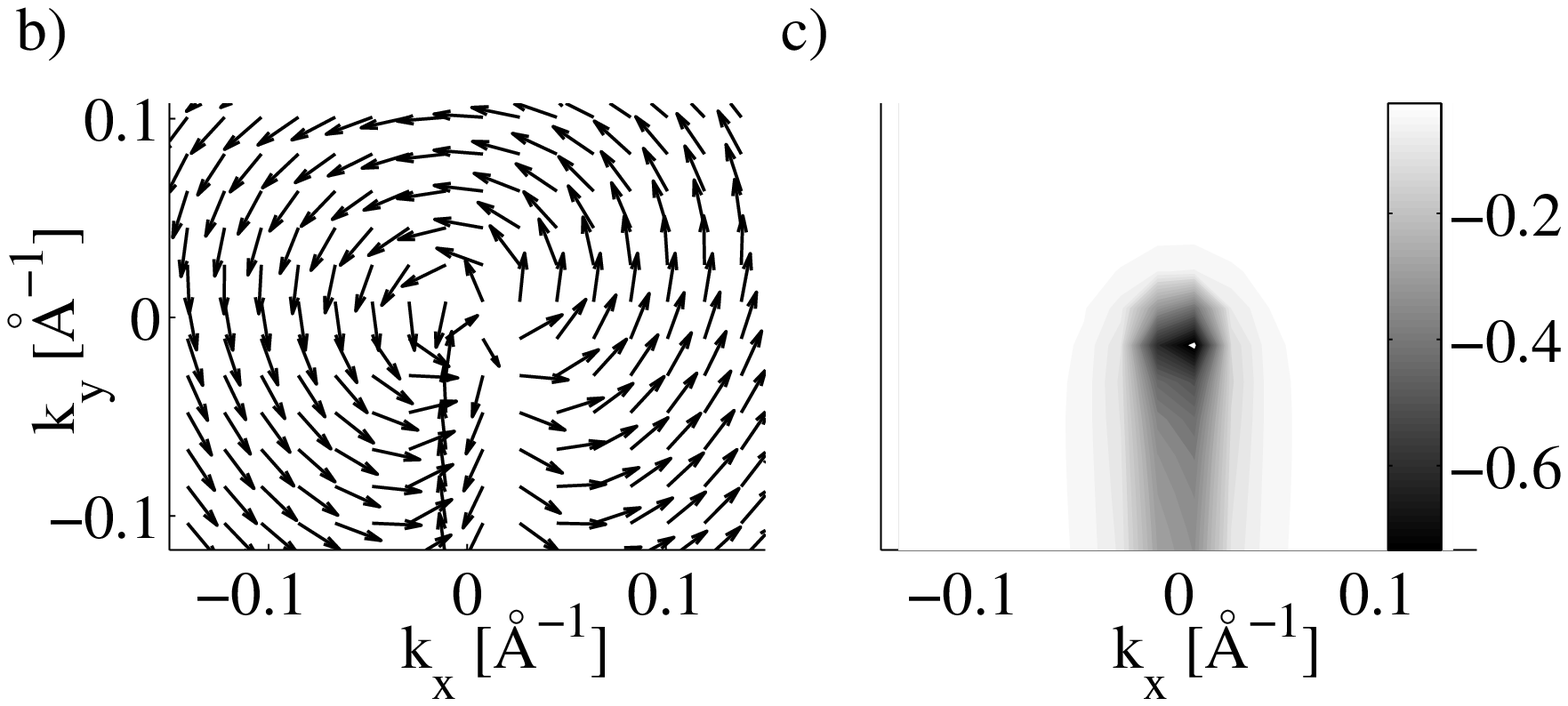}
\caption{ a) intensity distribution;   b) and c)  spin-polarization  of photoelectrons 
emanating from the  upper valance band ($\mu=-1$, $\nu=1$) for momentums close to the ${K}$ point. 
In b)  vector plot of the in-plane component of the spin polarization is shown. 
In c)  density plot of the $z$ component of spin  indicates the regions in the BZ where 
the out-of-plane polarization is finite.   
We used $\Delta=40\;{\rm meV}$, $\lambda_R=66\;{\rm meV}$ in these calculations.} 
\label{fig:spinpol}
\end{figure}

Using the explicit form of the quasiparticle spinors, calculations detailed 
in Appendix~\ref{sec:matrix_element} yield
\begin{equation}
%\begin{split}
  ^{\mu,\nu}I
 \propto %n_F(\varepsilon_{\mu\nu}({\bf k}))
\left(1 - \frac{v_F\hbar k_y (\mathcal{N}(k)-\nu\lambda_R^2)}{\mathcal{N}(k)\varepsilon_{\mu\nu}({\bf k})}\right).
%  &\delta_{\bf p_{\parallel}/\hbar - k - G,0}\;
%\delta (\hbar\omega + \varepsilon_{\mu\nu}({\bf k})-E_{{\bf p'},\sigma}-A)\;, 
\label{eq:ARPESint}
 %\end{split}
\end{equation}
As in previous theoretical works\cite{falko,kuemmeth,shirley} 
(which however did not consider either RSOI\cite{shirley,falko} or sublattice asymmetry\cite{kuemmeth}) 
we find a strongly anisotropic photoelectron intensity [see e.g. numerical results in Fig.~\ref{fig:spinpol}(a)] 
which originates from  sublattice interference\cite{falko} and therefore it is present\cite{intensity-pattern}
even if $\lambda_R=0$. 
Such anisotropy was  observed experimentally\cite{varykhalov,giant,zhou,bostwick-1,bostwick-1,sic-gap} too.
Indeed, Eq.~(\ref{eq:ARPESint}) for large wave numbers ($\hbar v_F k\gg |\lambda_R|$)  
can be approximated by
\begin{equation}
 %\begin{split}
^{\mu,\nu}I 
\propto 
\left(1 - \frac{v_F\hbar k_y }{\varepsilon_{\mu\nu}({\bf k})}\right) %\\
%&\delta_{\bf p_{\parallel}/\hbar - k - G,0}\;\delta 
%(\hbar\omega + \varepsilon_{\mu\nu}({\bf k})-E_{{\bf p'},\sigma}-A)\;, 
% \end{split} 
\label{eq:SARPESintapprox}
\end{equation}
from where it is easy to see that the intensity is minimal in the region where 
$k_y\gg |k_x|,|\Delta|/(\hbar v_F),|\lambda_R|/(\hbar v_F)$ \emph{and} $\mu k_y>0$. 
(By introducing the parametrization $(k_x,k_y)=k\,(\sin\theta, \cos\theta)$ one can see 
that $^{\mu,\nu} I$ takes a similar form to the result of Ref.~\onlinecite{kuemmeth}, though in our 
notation the indices ${\mu,\nu}$ have slightly different meaning.) 
Since the intensity of photoelectrons tends to vanish in this region, 
the authors of Refs.~\onlinecite{giant,kuemmeth} called this region  a \emph{dark corridor}. 
For  $\mathbf{k}\rightarrow 0$, in contrast, the intensity is isotropic. 
[We note that as the recent experiment of Gierz \emph{et al} showed 
[Ref.~(\onlinecite{s-polar})],  the angular distribution of photoelectron intensity 
also depends on the energy and polarization of the incident light. 
Our calculations should be relevant  for  $p$-polarized incident light.]

In terms of $\widetilde{Q}^{\mu,\nu}$, the expectation value of  an operator $\hat{O}$  which gives the result 
of a measurement on photoelectrons coming from band $(\mu,\nu)$ 
is given by
\begin{equation}
%\begin{split}
 ^{\mu,\nu}\langle O \rangle({\bf p}) = \bigg.{\rm Tr}\;
\bigg(\hat{O}\widetilde{Q}^{\mu,\nu}({\bf k})\bigg)  \;\bigg/\; {\rm Tr}\;
\bigg(\widetilde{Q}^{\mu,\nu}({\bf k})\bigg). \\
%&\delta_{ \mathbf{p}_{\parallel}/\hbar-(\mathbf{K}+\mathbf{k}+\mathbf{G}),\mathbf{0}}\;\delta 
%(\hbar\omega + \varepsilon_{\mu\nu}(\mathbf{k})-E_{{\bf p},\sigma}-W)\;. 
\label{eq:O_Tr}
%\end{split}
\end{equation}

We  make use of Eq.~(\ref{eq:O_Tr}) to calculate the photoelectron  spin-polarization vector
$(\mathcal{P}_x,\mathcal{P}_y,\mathcal{P}_z)=\frac{2}{\hbar}
(\langle \hat{s}_x\rangle, \langle \hat{s}_y\rangle, \langle \hat{s}_z\rangle)$
where the operator $\hat{s}_{x,y,z}=\sigma_{x,y,z}$ acts on the photoelectron spin.   
Using Eqs.~(\ref{eq:O_Tr}) and (\ref{eq:ARPESint}) we find for the components of the
polarization that  
\begin{subequations}
\begin{equation}
%\begin{split}
 ^{\mu,\nu}\mathcal{P}_x
\propto 
\frac{\nu\lambda_R\left(\frac{3}{2}\lambda_R^2 + 
\frac{2}{3}v_F\hbar k_y\left(v_F\hbar k_y - \varepsilon_{\mu\nu}({\bf k}) \right) \right) - 
\frac{3}{2}\lambda_R\mathcal{N}(k)} {\mathcal{N}(k)\varepsilon_{\mu\nu}({\bf k})-v_F\hbar k_y 
(\mathcal{N}(k)-\nu\lambda_R^2)}, %\\
%& \delta_{\bf p_{\parallel}/\hbar - k - G,0}\;\delta (\hbar\omega + 
%\varepsilon_{\mu\nu}({\bf k})-E_{{\bf p'},\sigma}-A)\;,
%\end{split} 
\label{eq:SARPESx} 
\end{equation}
\begin{equation}
%&\begin{split}
 ^{\mu,\nu}\mathcal{P}_y
\propto \nu\; \frac{\frac{2}{3}v_F\hbar k_x\;
\lambda_R\left( \varepsilon_{\mu\nu}({\bf k}) - v_F\hbar k_y \right)}
{\mathcal{N}(k)\varepsilon_{\mu\nu}({\bf k})-v_F\hbar k_y (\mathcal{N}(k)-\nu\lambda_R^2)},  %\\
%& \delta_{\bf p_{\parallel}/\hbar - k - G,0}\;\delta (\hbar\omega + \varepsilon_{\mu\nu}({\bf k})-
%E_{{\bf p'},\sigma}-A)\;,\end{split} 
\label{eq:SARPESy} 
\end{equation}
and
\begin{equation}
%&\begin{split}
 ^{\mu,\nu}\mathcal{P}_z
\propto \nu\;\frac{\frac{1}{2}\Delta \lambda_R^2}{\mathcal{N}(k)
\varepsilon_{\mu\nu}({\bf k})-v_F\hbar k_y (\mathcal{N}(k)-\nu\lambda_R^2)}. %\\
%& \delta_{\bf p_{\parallel}/\hbar - k - G,0}\;
%\delta (\hbar\omega + \varepsilon_{\mu\nu}({\bf k})-E_{{\bf p'},\sigma}-A)\;.
%\end{split}
\label{eq:SARPESz}
\end{equation} 
\label{eq:SARPES-photoe}
\end{subequations}

It is immediately clear from Eq.~(\ref{eq:SARPESz}) that similarly to Bloch electrons, 
photoelectrons also
acquire a finite $z$ polarization due to the interplay of sublattice asymmetry and RSOI. 
The magnitude of $^{\mu,\nu}\mathcal{P}_z$ is largest at the Dirac point for the $\nu=1$ bands where it 
reaches the value of unity. For the $\nu=-1$ bands the photoelectron polarization is
smaller:  $^{\mu,-1}\mathcal{P}_z(k=0)=-\Delta/\sqrt{\Delta^2+36\lambda_R^2}$.
In fact, as the density plot in Fig.~\ref{fig:spinpol}(c) shows for the upper valence band, 
$^{-1,1}\mathcal{P}_z(\mathbf{k})$ is finite everywhere in the dark corridor and is very small outside it 
(the plots for other bands are  similar and thus not shown). 
This suggests that in a constant energy SARPES measurement 
the easiest way to observe the finite $z$ polarization is to use energies close to
the Dirac point, otherwise one would have to collect data from the dark corridor, which is difficult 
due to the low photoelectron intensity and spin-detector efficiency.

Regarding the in-plane component of the photoelectron spin, Ref.~\onlinecite{kuemmeth} has found 
that in the case of equivalent sublattices  it exhibits a rather peculiar behavior, 
especially in and close to the dark corridor, where the 
photoelectron spin is rotated with respect to the quasiparticle spin. 
Moreover, Ref.~\onlinecite{kuemmeth} also showed that the in-plane spin polarization 
of photoelectrons is not zero in the ${K}$ point even though the mean spin of 
Bloch electrons is zero there [see Eqs.~(\ref{eq:spin-xy})].  
We find from Eqs.~(\ref{eq:SARPESx}) and (\ref{eq:SARPESy}) that the breaking of the $AB$ 
symmetry does not alter significantly this picture of the in-plane polarization, 
thus we will only briefly discuss it. An example of the 
photoelectron in-plane spin polarization is shown in Fig.~\ref{fig:spinpol}(b) for the 
upper valence band. One can clearly observe that the spins are rotated in the dark corridor 
(at $k_y<0$, $k_x\approx 0$,  see   Fig.~\ref{fig:spinpol}(a) where the intensity map is 
shown for the same band).  In contrast to the in-plane spin of quasiparticles, the corresponding
spin component of photoelectrons therefore does not show   rotational symmetry around the ${K}$ point.

The opening of a small gap at the Dirac point due to the $AB$ symmetry breaking effect of a  substrate 
is not easy to detect in an ARPES measurements  because of the  finite energy resolution of the experiments and 
because of the energy broadening of the bands. In the next section we investigate the possibility of 
detecting the sublattice asymmetry through photoelectron spin polarization. 
To this end we compute the intensity 
maps  and spin polarization distributions of photoelectrons at given energies.

\section{Numerical (S)ARPES calculations}
\label{sec:num-sarpes}

In this section we discuss the results of numerical calculations of  constant-energy intensity 
maps and spin polarizations along certain directions in the BZ. In Ref.~\onlinecite{sajat} we showed
that the Hamiltonian of monolayer graphene for finite RSOI  is formally the same as the Hamiltonian 
of bilayer graphene, if certain weak hopping amplitudes in the latter system can be 
neglected.  The aim of this section is twofold. Firstly, we  point out both the similarities 
and the differences in  the constant energy ARPES intensity maps of monolayer graphene with 
RSOI and bilayer graphene. Secondly, we  show  photoelectron spin polarization calculations along 
certain directions in the BZ (spin-resolved MDCs) and relate them to the fixed energy ARPES intensity maps.
In the calculation of spin-resolved MDCs we assume that the background is small and 
disregard its influence on the lineshapes\cite{sarpes-1}.    
Since the $AB$ asymmetry does not break the 
particle-hole symmetry of the Hamiltonian, we only show calculations for energies in the valence bands. 
We assume strong RSOI and use $\lambda_R=66\,{\rm meV}$ corresponding to $\approx200\,{\rm meV}$
 spin-splitting of the bands\cite{giant}.

We start the discussion with 
intensity maps taken at energies close to the Dirac point. 
In the derivation of Eq.~(\ref{eq:ARPESint}) we neglected those terms in the 
graphene Hamiltonian which cause trigonal warping of the bands for low energies if RSOI is finite 
[see the discussion below Eq.~(\ref{eq:H_RSO})]. This approximation is useful  
to understand the main features of the  spin-polarization  but for strong RSOI the neglected 
terms do cause a noticeable change in the fixed-energy intensity maps. In the calculations shown 
below therefore we take   these terms into account as well.

\begin{figure}
\includegraphics[width=0.49\textwidth]{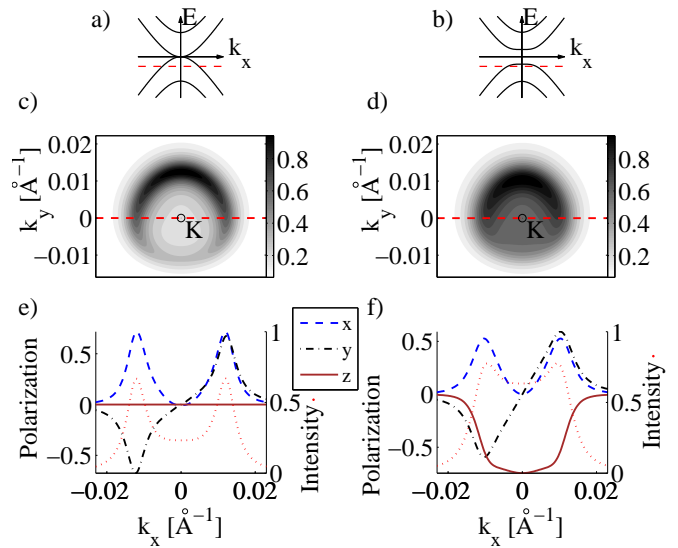}
\caption{ Constant energy (S)ARPES calculations close to the Dirac point.    
 a) and b): schematic band structure at the ${K}$ point of the BZ for zero and 
finite sublattice asymmetry, respectively. 
Dashed lines indicate a constant energy cross-section at $E=-37\,{\rm meV}$ where the photoelectron
intensity maps in c) and d) are obtained. 
c) and  d): constant energy intensity maps for $\Delta=0$ and $\Delta=40\,{\rm meV}$, respectively.
The dashed lines  indicate the direction 
in $\mathbf{k}$ space along which the spin  polarization curves in  e) and  f)  are calculated. 
e) and f): the $x$ (dashed), $y$ (dashed-dotted) and $z$ (solid) component of the 
photoelectron spin polarization for  cross sections  shown in c) and d), respectively.
(left axis). The dotted lines indicate the intensity profile along the same cross-section 
(right axis). 
Subfigures in the left [right] column correspond to sublattice anisotropy parameter 
$\Delta=0$ [$\Delta=40\;{\rm meV}$].
Other parameters of the figure are $\lambda=66\,{\rm meV}$ and $\Gamma=12.5\,{\rm meV}$.} 
\label{fig:SARPES-lowenergy}
\end{figure}

\begin{figure}
\includegraphics[width=0.49\textwidth]{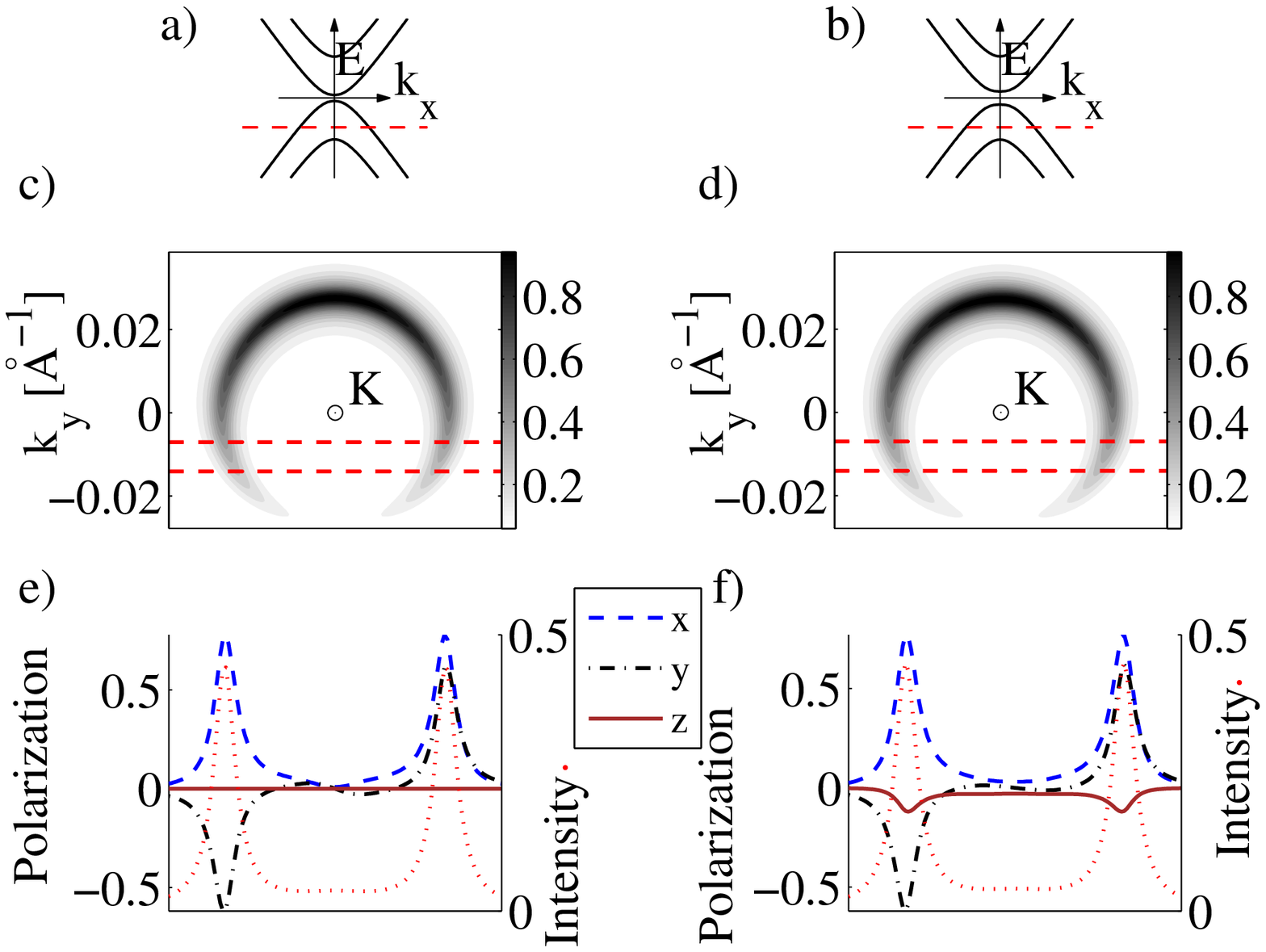}
\includegraphics[width=0.49\textwidth]{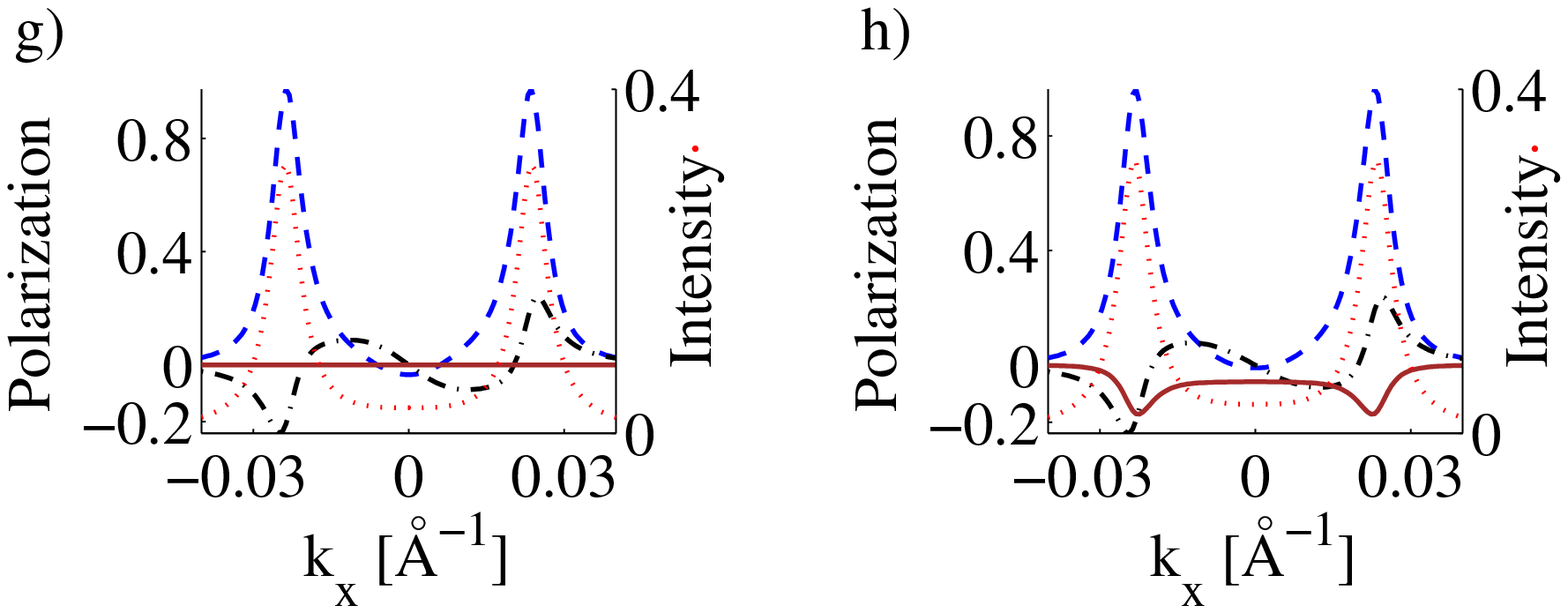}
\caption{ Constant energy (S)ARPES calculations for low energies and small broadening  $\Gamma$.
a) and b): schematic band structure at the ${K}$ point of the BZ 
for zero and finite sublattice asymmetry, respectively. 
Dashed lines indicate a constant energy cross-section at $E=-110\;{\rm meV}$ where 
the photoelectron intensity maps in c) and d) are obtained. 
c) and d): constant energy intensity maps for $\Delta=0$ and $\Delta=40\;{\rm meV}$, respectively. 
The dashed line at $k_y=-0.007 {\rm \AA^{-1}}$ [$k_y=-0.014 {\rm \AA^{-1}}$] indicate the direction 
in $\mathbf{k}$ space along which the spin  polarization curves in subfigures  
e) and f)  [ g) and h) ] are calculated. 
e) and g) [ f) and h) ]: the $x$ (dashed), $y$ (dashed-dotted) and $z$ (solid) component of the 
photoelectron spin polarization for the two cross sections  shown in c) [ d)] (left axis). 
The dotted lines indicate the  intensity profile along the same cross-section (right axis). 
Subfigures in the left [right] column correspond to sublattice anisotropy parameter 
$\Delta=0$ [$\Delta=40\;{\rm meV}$].
Other parameters of the figure are $\lambda=66\;{\rm meV}$ and $\Gamma=16.7\;{\rm meV}$.} 
\label{fig:SARPES-midenergy}
\end{figure}

In Figs.~\ref{fig:SARPES-lowenergy}(c) and \ref{fig:SARPES-lowenergy}(d) 
only a small  broadening of the lines is assumed. Because
of the strong RSOI ($\lambda_R=66\,{\rm meV}$), small $\Gamma$  and low energy ($E=-37\rm{meV}$) 
the photoelectrons come predominantly 
from the upper valence band. 
One can observe the following important features:  
similarly to monolayer graphene
with zero RSOI (Refs.~\onlinecite{shirley,falko}) there is  a characteristic angular variation 
in the intensity which is  due to   sublattice interference and particularly 
 in Fig.~\ref{fig:SARPES-lowenergy}(c) one can see  a low intensity region 
(the ''dark corridor``) around $k_x\approx 0$ and $k_y<0$. Nevertheless, as a consequence of 
spin-pseudospin entanglement\cite{kuemmeth} at these low energies 
the intensity distribution is  more isotropic in the case of finite RSOI 
than it is  for $\lambda_R=0$. 
Fig.~\ref{fig:SARPES-lowenergy}(d) shows that the main effect of finite sublattice asymmetry 
on the intensity maps is that  it  reduces the intensity anisotropy clearly seen in
 Fig.~\ref{fig:SARPES-lowenergy}(c). Comparing e.g.  Fig.~\ref{fig:SARPES-lowenergy}(c) and 
 Fig.~\ref{fig:SARPES-midenergy}(c) one can also notice that in the former figure there is 
a slight trigonal distortion in the intensity contour. This  distortion, which is
caused by the terms $v_{\lambda}\hat{p}_{\pm}$ in the Hamiltonian (\ref{eq:H_RSO}), 
 can only be seen for strong RSOI and close to the charge neutrality point. Note, that it is
different  from the trigonal distortion observable 
 for energies far from the Dirac point (see Fig.~\ref{fig:SARPES-highenergy}) which is a lattice effect.

Figs.~\ref{fig:SARPES-lowenergy}(e) and \ref{fig:SARPES-lowenergy}(f) show the spin polarization 
as a function of momentum along the direction  indicated by dashed line in 
Figs. ~\ref{fig:SARPES-lowenergy}(c) and \ref{fig:SARPES-lowenergy}(d), 
respectively. As evidenced by the peaks in the $x$ polarization component $\mathcal{P}_x$
[and also noted in Ref.~\onlinecite{kuemmeth}], in contrast to Bloch electrons,
the spin polarization of  photoelectrons is not necesseraly transversal to the momentum $\mathbf{k}$.   
 One can also see  that $\mathcal{P}_y$ changes sign as  the $k_x=0$ line is crossed. 
The  out-of-plane component of the photoelectron spin is zero if no sublattice asymmetry 
is assumed [Fig.~\ref{fig:SARPES-lowenergy}(f)] but  $\mathcal{P}_z$ is  
finite if $\Delta\neq 0$, as in Fig.~\ref{fig:SARPES-lowenergy}(e). This means that 
through  SARPES measurements in systems where RSOI is nonzero  the $AB$ asymmetry can be detected 
\emph{even if the sample is slightly $p$-doped}, i.e. states around the Dirac-point are not 
directly available by ARPES.

In Fig.~\ref{fig:SARPES-midenergy} the constant energy maps are 
calculated at $E=-110\,{\rm meV}$, i.e. not in the close vicinity of the charge neutrality point. 
As the schematic figures ~\ref{fig:SARPES-midenergy}(a) and ~\ref{fig:SARPES-midenergy}(b) show,
because of the large spin-splitting (and a small broadening of $\Gamma=16.7\;{\rm meV}$) 
assumed, all the photoelectrons would still originate from the same  band as in the previous case.  
The intensity maps in Figs.~\ref{fig:SARPES-midenergy}(c) and (d) resemble closely the corresponding maps of 
  monolayer graphene (see e.g. Fig.~2 in Ref.~\onlinecite{falko}). In particular, one can observe 
 an almost complete suppression of intensity in the dark corridor and the disappearance of 
the trigonal distortion of the intensity maps, apparent in Figs.~\ref{fig:SARPES-lowenergy}(c) and (d).
Furthermore, comparing Fig.~\ref{fig:SARPES-midenergy}(c) and  Fig.~\ref{fig:SARPES-midenergy}(d) one 
can see that the presence of a small asymmetry gap 
[$\Delta=40\,{\rm meV}$ in Fig.~\ref{fig:SARPES-midenergy}(d)] would be practically  
undetectable in an ARPES measurement at this energy. Nevertheless, as 
Fig.~\ref{fig:SARPES-midenergy}(f) and Fig.~\ref{fig:SARPES-midenergy}(h) show,  
if $\Delta\neq 0$ there is a small but finite $z$ polarization component. 
Comparison of  Fig.~\ref{fig:SARPES-midenergy}(f) and Fig.~\ref{fig:SARPES-midenergy}(h) 
illustrates the feature shown in Fig.~\ref{fig:spinpol}(c): $\mathcal{P}_z(\mathbf{k})$ is 
largest in the dark corridor, therefore in a constant energy measurement it is larger if 
the direction in the $\mathbf{k}$ space is chosen such that it is closer to the dark corridor 
[ Fig.~\ref{fig:SARPES-midenergy}(f) is calculated for $k_y=-0.007 {\rm \AA^{-1}}$ with 
maximal polarization of $\mathcal{P}_z^{max}=-0.12$, whereas   $k_y=-0.14 {\rm \AA^{-1}}$ 
in Fig.~\ref{fig:SARPES-midenergy}(h) and $\mathcal{P}_z^{max}=-0.17$]. Note, that 
even in the case of  Fig.~\ref{fig:SARPES-midenergy}(h) the curve is not actually 
calculated in the dark corridor, the ARPES intensity peaks (shown by dotted line) 
for this cross-section are roughly $40\%$ of the maximum intensity that can be found 
at this energy [black arc close to the upper edge of Fig.~\ref{fig:SARPES-midenergy}(d)]. 
On the other hand, the in-plane components of the spin polarization are practically the same for the 
$\Delta=0$ [Figs.~\ref{fig:SARPES-midenergy}(e), \ref{fig:SARPES-midenergy}(g)] and 
$\Delta\neq 0$ [Figs.~\ref{fig:SARPES-midenergy}(f), \ref{fig:SARPES-midenergy}(h)] cases.

\begin{figure}
\includegraphics[width=0.49\textwidth]{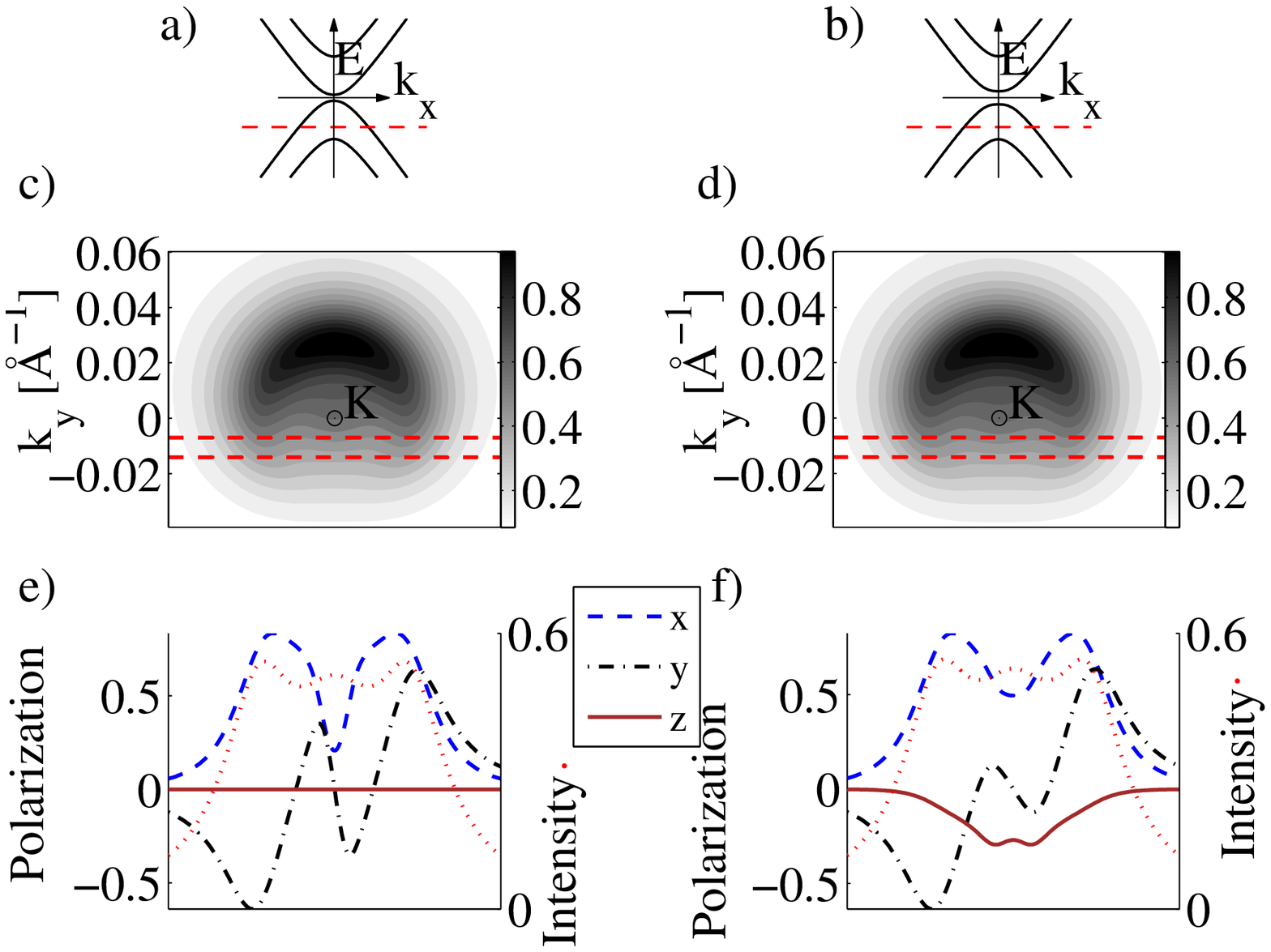}
\includegraphics[width=0.49\textwidth]{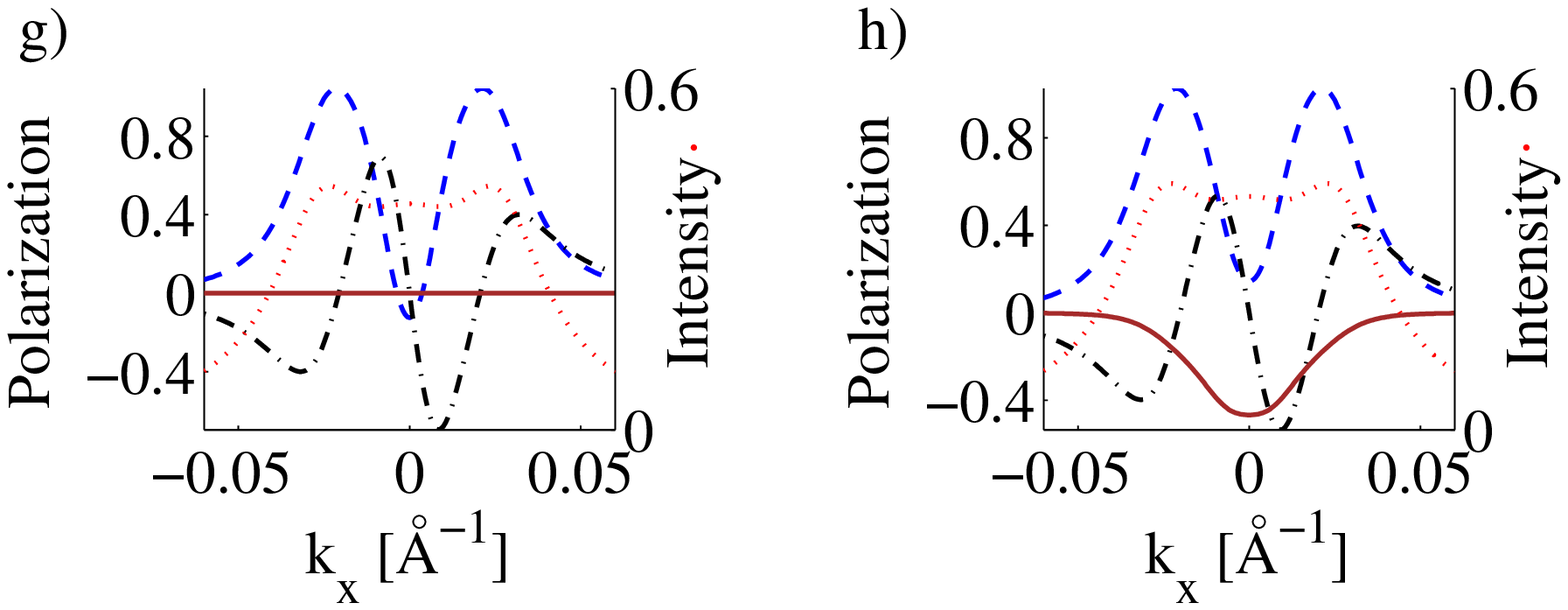}
\caption{ The effects of the broadening parameter $\Gamma$ on the (S)ARPES spectra. 
As in Fig.~\ref{fig:SARPES-midenergy}, all calculations are for $E=-110\,{\rm meV}$.
a) and b): the same as in Fig.~\ref{fig:SARPES-midenergy}.  
c) and d): constant energy intensity maps for $\Delta=0$ and $\Delta=40\;{\rm meV}$, respectively.
The dashed line  at $k_y=-0.007 {\rm \AA^{-1}}$ [$k_y=-0.014 {\rm \AA^{-1}}$] indicate the direction 
in $\mathbf{k}$ space along which the spin  polarization curves in subfigures  e) and f) 
[g) and h)] are calculated. 
e) and g) [ f) and g) ]: the $x$ (dashed), $y$ (dashed-dotted) and $z$ (solid) component of the 
photoelectron spin polarization for the two cross-sections  shown in c) [ d) ] (left axis). 
The dotted lines indicate the  intensity profile along the same cross-section (right axis). 
Subfigures in the left [right] column correspond to sublattice anisotropy parameter 
$\Delta=0$ [$\Delta=40\;{\rm meV}$].
Other physical parameters of the figure are $\lambda=66\;{\rm meV}$ and $\Gamma=83.5\;{\rm meV}$.} 
\label{fig:SARPES-largeGamma}
\end{figure}
In ARPES measurements the energy broadening is often quite substantial. To see the 
effects of broadening on the SARPES spectra  we repeated the 
calculations shown in Fig.~\ref{fig:SARPES-midenergy} for a larger broadening parameter. 
The results for  $\Gamma=83.5\;{\rm meV}$ are presented  in Fig.~\ref{fig:SARPES-largeGamma}. 
Although the ARPES fixed energy contours are significantly blurred due to the large  $\Gamma$, 
[Figs.~\ref{fig:SARPES-largeGamma}(c) and \ref{fig:SARPES-largeGamma}(d)] the broadening  
would actually lead to a bigger  out-of-plane spin polarization amplitude 
see Figs.~\ref{fig:SARPES-largeGamma}(f) and \ref{fig:SARPES-largeGamma}(h)], 
hence it would make the detection of 
the $z$ spin polarization easier [c.f  Figs.~\ref{fig:SARPES-midenergy}(f) and \ref{fig:SARPES-midenergy}(h)]. 
This happens because for large broadening electrons having energies closer to the Dirac 
point can also contribute and they have larger spin $z$ component.   
Other noticeable feature in Figs.~\ref{fig:SARPES-largeGamma}(e)-\ref{fig:SARPES-largeGamma}(h) 
compared to Figs.~\ref{fig:SARPES-midenergy}(e)-\ref{fig:SARPES-midenergy}(h)
is that one can clearly see that $\mathcal{P}_y$ changes sign three times for small $k_x$ values. 
This is not apparent in e.g. Figs.~\ref{fig:SARPES-midenergy}(e) and \ref{fig:SARPES-midenergy}(f) because 
of the small amplitude of these oscillations there.

Finally, we consider  the constant-energy intensity maps and spin polarizations  at  energy 
$E=-660\,{\rm meV}$, i.e. quite far from the Dirac point. For these calculations we used the 
tight-binding Hamiltonian of Ref.~\onlinecite{kane}. 
Since this energy is larger than 
the spin splitting $3\lambda_R=200\,{\rm meV}$ used in our calculations,  
both valence bands contribute to  the ARPES and  SARPES spectra. We assume for simplicity 
that the broadening $\Gamma$ is the same for both bands and present calculations
with  two different  $\Gamma$s,
the first one being much smaller than the spin-splitting of the bands, while the second one is 
comparable to it. 
\begin{figure}
\includegraphics[width=0.49\textwidth]{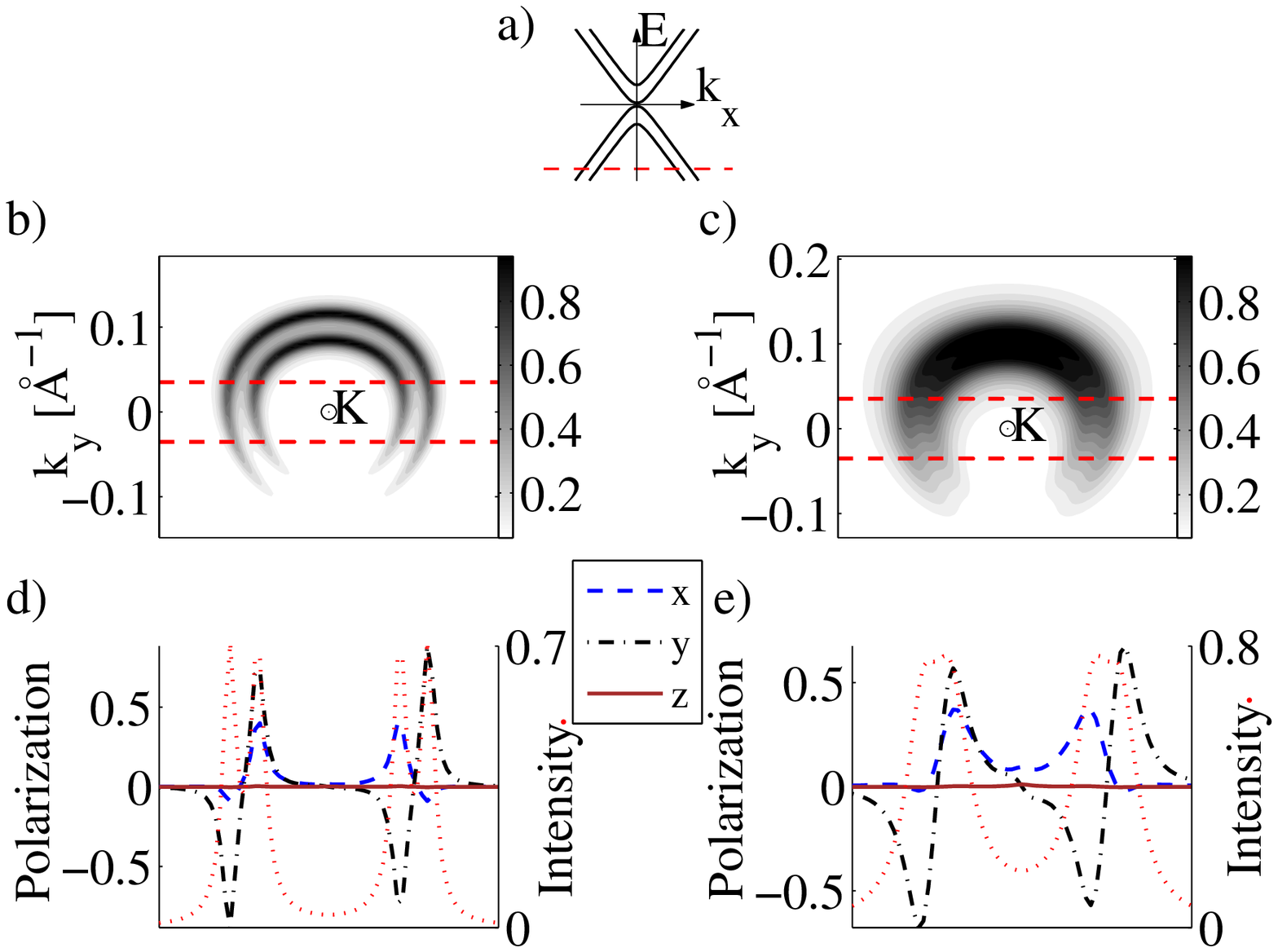}
\includegraphics[width=0.49\textwidth]{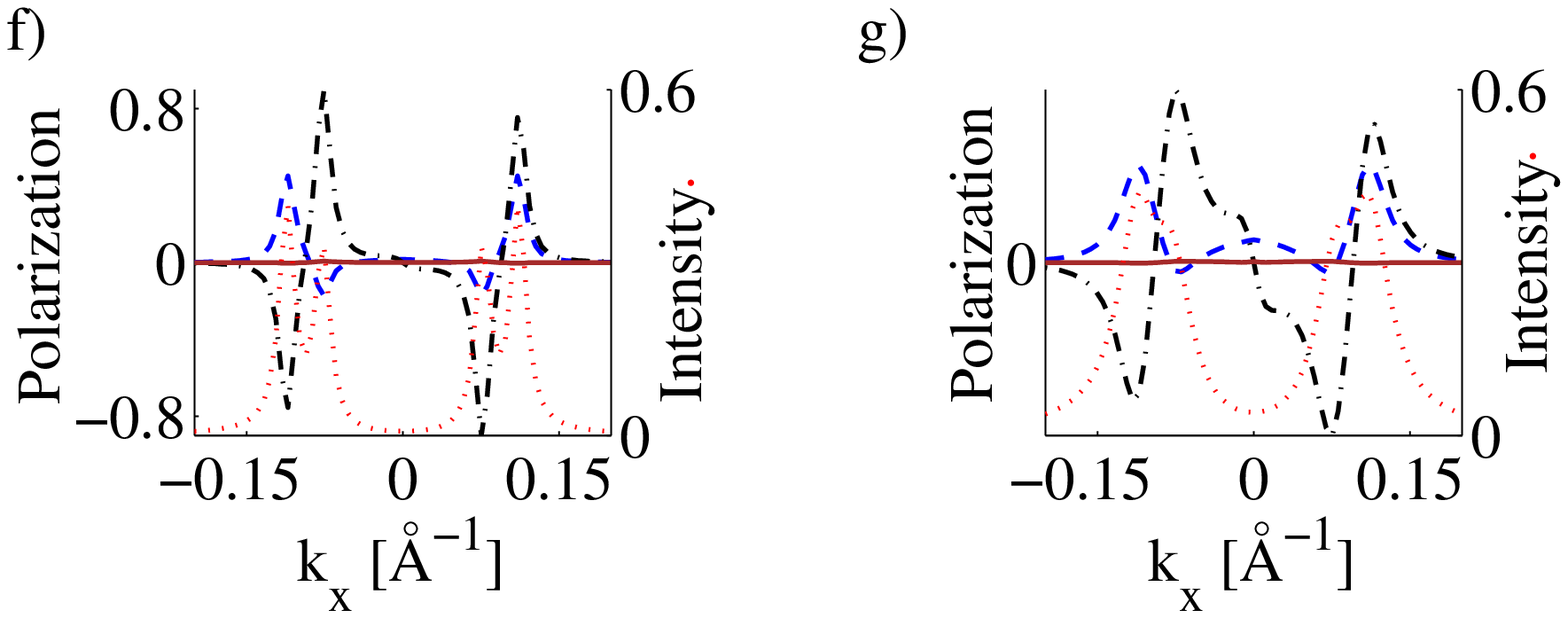}
\caption{ Constant energy (S)ARPES calculations far from the Dirac point.
a): schematic band structure at the ${K}$ point. 
Dashed line indicates a constant energy cross-section at $E=-660\;{\rm meV}$ where the photoelectron
intensity maps in b) and c) are obtained. 
b) and c): constant energy intensity maps for $\Gamma=50\;{\rm meV}$ and $\Gamma=134\;{\rm meV}$, 
respectively. The dashed lines at $k_y=0.035 {\rm \AA^{-1}}$  [$k_y=-0.035 {\rm \AA^{-1}}$] 
indicate the direction 
in $\mathbf{k}$ space along which the spin  polarization curves in subfigure  d) and e) 
[ f) and g) ] are calculated. 
d) and f) [e) and g)]: the $x$ (dashed), $y$ (dashed-dotted) and $z$ (solid) component of the 
photoelectron spin polarization for the two cross sections  shown in b) [ c) ] (left axis). 
The dotted lines indicate the  intensity profile along the same cross-section (right axis).  
\label{fig:SARPES-highenergy}}
\end{figure}
As  $E\gg\Delta$ in this case, the ARPES and SARPES spectra are practically the same for
$\Delta=0$ or $\Delta\neq 0$,  therefore  we only show results for $\Delta=0$.

If the broadening is moderate, as in Fig.~\ref{fig:SARPES-highenergy}(b), there are two discernible ringlike 
patterns, each corresponding to photoemission from states in one of the two bands. 
The rings show slight trigonal distortion, but in contrast to Fig.~\ref{fig:SARPES-lowenergy}(c),  
this  is a lattice effect and would also be observable\cite{falko} for $\lambda_R=0$. 
%(we are far in energy from the charge neutrality point where the dispersion is given by a simple Dirac cone).
The double ringlike pattern is reminiscent of the intensity maps found for 
bilayer graphene at high energies\cite{falko,bilayer-arpes}, but an important difference is that in 
Fig.~\ref{fig:SARPES-highenergy}(b) both rings have approximately the same intensity. 
The similarities between the ARPES maps of the two systems are due to the similar bandstructures  
(for a discussion of the relation between the Hamiltonians of monolayer graphene with RSOI and 
bilayer graphene see Ref.~\onlinecite{sajat}). 
The difference in the intensity patterns stems from the fact that there are four carbon atoms in the 
unit cell of bilayer while there are only two in monolayer graphene therefore the 
transition matrix elements in the photoemission calculations are different. 

If the broadening is substantial, as in  Fig.~\ref{fig:SARPES-highenergy}(c) , the two rings 
are no longer easily discernible (and they may even completely overlap). Nevertheless, as 
the dashed-dotted curves in Figs.~\ref{fig:SARPES-highenergy}(e) and \ref{fig:SARPES-highenergy}(g) 
demonstrate, the $y$ component of the spin polarization changes sign as a function of $k_x$ roughly in the 
middle of the intensity peak (dotted line). This  is 
an indication that two bands are involved in the photoemission, as  the sign of $^{\mu,\nu}\mathcal{P}_y$
is different for the $\nu=1$ and $\nu=-1$ bands [see Eq.~(\ref{eq:SARPESy})].  Furthermore, comparison 
of Figs.~\ref{fig:SARPES-highenergy}(e) and \ref{fig:SARPES-highenergy}(g) 
[Figs.~\ref{fig:SARPES-highenergy}(d) and \ref{fig:SARPES-highenergy}(f)] shows
that the overall shape and the number of sign changes in $^{\mu,\nu}\mathcal{P}_y$ do not depend on whether 
it is calculated for a positive or negative $k_y=const$ value 
[see Fig.~\ref{fig:SARPES-highenergy}(b) or \ref{fig:SARPES-highenergy}(c) for the cross-sections
along which Figs.~\ref{fig:SARPES-highenergy}(d)-\ref{fig:SARPES-highenergy}(g) were obtained]. 
In contrast, for $^{\mu,\nu}\mathcal{P}_x$ the number of sign changes in the low intensity 
region (small $|k_x|$ values) is affected by the choice of the $k_y$, as e.g the comparison of 
Figs.~\ref{fig:SARPES-highenergy}(e) and  \ref{fig:SARPES-highenergy}(g) can illustrate.

\section{Discussion and Summary}
\label{sec:summary}

We would first briefly comment on the experimental relevance of our results.  
As mentioned in the Introduction, a significant spin-orbit coupling was found in gold intercalated 
Ni(111)/graphene system\cite{varykhalov} and the SOI was attributed to the presence of the 
gold atoms. Spin-resolved MDCs were not shown however in Ref.~\onlinecite{varykhalov}. 
Subsequently,  Ref.\onlinecite{ruthenium-gap} proved 
that gold intercalation can decouple  graphene from the Ru(0001) surface as well. 
Another notable recent development is that gold intercalation has also been  used for the 
 Si face of SiC substrate\cite{gierz-gold} where the strong covalent bonds between the 
 SiC(0001) and the first graphitic layer were suppressed by this method resulting 
 in a slightly p-doped graphene that was only weakly influenced by the substrate.
SARPES measurements were not published in Refs.~\onlinecite{ruthenium-gap,gierz-gold}, though 
it would be interesting to know if gold can induce SOI in these systems as well.    
The    Ru(0001)/gold/graphene system appears to be particularly interesting from 
our point of view because ARPES measurement indicate  a band gap $> 100\,{\rm meV}$, so that 
if RSOI is non-zero  in this system then a finite out-of-plane polarization
of photoelectrons should be measurable. 
A qualitatively similar polarization  to the one predicted by this model, with an 
''abrupt rotation of the spin`` at the $K$ point of the BZ was measured  
when thallium was deposited on Si(111) surface\cite{sakamoto}, though 
Ref.~\onlinecite{sakamoto} explained the effect by the presence of a local effective magnetic field.   
Finally, we note that  Ref.~\onlinecite{giant} reported a large and anisotropic spin splitting  
in graphene, including a nonzero out-of plane polarization component, but the origin 
of this effect seems to be unclear at the moment.

In summary, we studied the effect of RSOI and substrate induced sublattice asymmetry on the spin 
polarization of quasiparticles and of photoelectrons in graphene. 
The breaking of $AB$ sublattice symmetry opens a gap in the band structure of graphene at the 
$K$ point of the BZ. If RSOI is  finite, the interplay of the two effects induces a non-zero
out-of-plane component of spin polarization of quasiparticles in part of the BZ. 
RSOI also affects the intensity and spin distribution of photoelectrons, hence it can be studied 
with the (S)ARPES technique. 
For strong RSOI,  the fixed-energy intensity maps taken at low energies, 
close to the $K$ point of the BZ, 
show a characteristic trigonal deformation. This deformation of the intensity map  
survives the switching-on of an $AB$ symmetry breaking potential given by the asymmetry parameter 
$\Delta$, as long as   $\Delta$ is much smaller than the RSOI induced band splitting.       
Our spin-resolved MDCs calculations also show that an important sign of the simultaneous presence of  
RSOI and sublattice asymmetry is  if non-zero out-of-plane photoelectron 
spin polarization can be measured. It is important however, 
especially if $\Delta$ and RSOI are small, to choose the energy at which the spin-resolved MDCs are 
taken as close as possible to the Dirac-point, because for energies far from it the out-of-plane
polarization remains finite only in the ''dark corridor'', where the low 
photoelectron intensity would hinder the observation of this effect. 
A carefully chosen cross-section in the momentum space or  a large intrinsic energy 
broadening may, however, facilitate the observation of the spin $z$ polarization 
in MDCs even at higher energies. Meanwhile, the in-plane components of photoelectron polarization 
remain qualitatively the same regardless of whether $\Delta$ is zero or not.  
If the fixed-energy intensity map is obtained at  energies larger than the energy separation of 
two spin-split bands and their intrinsic energy broadening $\Gamma$  is small 
compared to their RSOI induced energy splitting, 
then the resulting ARPES calculation shows a double ring-like structure.  
For large $\Gamma$, the two rings may not be discernible any more, but  SARPES measurements can 
nevertheless reveal the  true band structure because of the sign-changes in the polarization
components.

\section{Acknowledgments}
This work was supported by the Marie Curie ITN project NanoCTM 
(FP7-PEOPLE-ITN-2008-234970),
the Hungarian Science Foundation OTKA under the contracts No. 75529 and 
No. 81492
and the European Union and the European Social Fund have provided 
financial support to the project
under the grant agreement  T\'AMOP 4.2.1./B-09/1/KMR-2010-0003. 
A.K. also acknowledges the support of EPSRC.

\appendix

\section{Outline  of the theoretical SARPES calculations} 
\label{sec:matrix_element}

Here we briefly describe the calculation leading to  Eq.~(\ref{eq:O_Tr}).   
The Hamiltonian of the interaction between the Bloch electrons  and the electromagnetic field
in dipole approximation\cite{shirley} is given by 
\begin{equation}
 \hat{H}_{int} \propto -\frac{\hbar}{{\rm i}} {\boldsymbol A\nabla}\;,
\end{equation}
where ${\bf A} = {\bf A_0}e^{{\rm i}({\bf qr}-\omega t)}$ is the vector potential.
The transition probability between an initial Bloch electron state 
$| {\bf k},(\mu,\nu)\rangle$ %of  momentum ${\bf K+k}$ in band $(\mu,\nu)$  
and a photoelectron state
$|{\bf p},\sigma\rangle$ will be proportional to 
$ 
|\left( H_{int} \right)_{\textbf{k},(\mu,\nu)}^{\textbf{p},\sigma}|^2\,\, 
\delta(\hbar\omega + \varepsilon_{\mu\nu}(\mathbf{k})-E_{{\bf p},\sigma}-W)
$, where the 
photoexcitation matrix element is  
\begin{equation}
\left( H_{int} \right)_{\textbf{k},(\mu,\nu)}^{\textbf{p},\sigma}= 
\langle{\bf p},\sigma |\hat{H}_{int}| {\bf k},(\mu,\nu)\rangle.
\label{eq:transit-matrx} 
\end{equation}
Explicit expression for $\left(H_{int}\right)_{\textbf{k},(\mu,\nu)}^{\textbf{p},\sigma}$ can obtained 
by assuming that  the wavefunction of a photoelectron given by a plane wave 
$|{\bf p},\sigma\rangle \propto 
e^{{\rm i}\bf pr/\hbar}|\sigma\rangle$ 
and the wavefunction of a Bloch electron is
\begin{equation}
\begin{split}
|{\bf k},(\mu,\nu)\rangle &= \frac{1}{\sqrt{\mathcal{N}({\bf k})}}
\sum\limits_{j = \{A,B\}, \sigma' = \{\uparrow,\downarrow\}}\Bigg[ \\
&\psi_{j\sigma'}^{\mu\nu}({\bf k}) \;|\sigma'\rangle\left[\frac{1}{\sqrt{N}} 
\sum\limits_{n=1}^Ne^{{\rm i}(\textbf{k+K}) \textbf{R}^j_n}\Phi({\bf r-R}^j_n) \right]\Bigg] \;.
\end{split} 
\label{eq:hullamfvBloch}
\end{equation}
Here ${\bf R}^j_n$ are vectors pointing to  sublattice sites $j=\{A,B \}$ in unit cell $n$, $N$ is the number of 
unit cells in the sample, $\psi_{j\sigma'}^{\mu\nu}({\bf k})$ are the 
amplitudes of Bloch electrons on sublattice $j$ with momentum ${\bf k}$ and spin $\sigma'$ and finally,
$\Phi(\mathbf{r})$ is a $p_z$ atomic orbital.  
The photoexcitation matrix element then reads:
\begin{equation}
%\begin{split}
\left( H_{int} \right)_{\textbf{k},(\mu,\nu)}^{\textbf{p},\sigma} \propto 
\sqrt{N} \Phi_p ({\bf A p}) \bigg(\psi_{A\sigma}^{\mu}({\bf k}) 
+ e^{{\rm i}{\bf G}{\boldsymbol\tau}} \psi_{B\sigma}^{\mu}({\bf k})\bigg). %\\
%\delta_{\bf p_{\parallel}/\hbar - (K+ k +G),0}\;. 
%\end{split}
\label{eq:Hint-explic} 
\end{equation}
In Eq.~(\ref{eq:Hint-explic}) $\Phi_{\mathbf{p}}$ is the Fourier transform  
of the atomic orbital $\Phi(\mathbf{r})$,  
and ${\bf G} = m_1{\bf b_1} + m_2{\bf b_2}$ is a reciprocal lattice vector which is given 
in terms of primitive lattice vectors  $\mathbf{b}_1=(2\pi/a_0,2\pi/\sqrt{3}\,a_0)$,  
$\mathbf{b}_2=(2\pi/a_0,-2\pi/\sqrt{3}\,a_0)$ and integers $m_1$, $m_2$. Furthermore, 
${\boldsymbol \tau} \equiv {\bf R^B_n - R^A_n}$  % $\frac{1}{3}\left({\bf a}_2 - {\bf a}_1 \right)$ 
 and ${\bf p}_{\parallel}$ is the  
projection of momentum $\bf p$ onto the plane of graphene. 
%It is easy to check that ${\bf G}{\boldsymbol\tau} = \frac{m_1+m_2}{3}\;.$
By defining 
\begin{equation}
 |\Phi_{\bf p}^{(\mu,\nu)}\rangle = \sum_{\sigma=\{\uparrow,\downarrow\}}
\left( H_{int} \right)_{\textbf{k},(\mu,\nu)}^{\textbf{p},\sigma}|{\bf p},\sigma\rangle, 
\end{equation}
the  expectation value of an operator $\hat{O}$ with respect to the 
photoelectron state emanating from an  initial  Bloch state of momentum $\hbar(\mathbf{K}+\mathbf{k})$, 
energy $\varepsilon_{\mu\nu}(\mathbf{k})$ in band $(\mu,\nu)$ can be calculated as 
 \begin{equation}
\begin{split}
 ^{\mu,\nu}\langle O \rangle({\bf p}) = &\frac{\langle\Phi_{\bf p}^{(\mu,\nu)}|\hat{O}|\Phi_{\bf p}^{(\mu,\nu)}\rangle}
{\langle\Phi_{\bf p}^{(\mu,\nu)}|\Phi_{\bf p}^{(\mu,\nu)}\rangle} \times \\
&\delta_{\bf p_{\parallel}/\hbar - (K+ k +G),0}\,
\delta(\hbar\omega + \varepsilon_{\mu\nu}(\mathbf{k})-E_{{\bf p},\sigma}-W).
\end{split} 
\label{eq:expect-val-O}
\end{equation}
Eq.~(\ref{eq:O_Tr}) then follows from  Eqs.~(\ref{eq:Hint-explic}) and (\ref{eq:expect-val-O}). 
We note that a convenient way of calculating the projectors $Q^{\mu, \nu}({\bf k})$ which are necessary
to evaluate Eqs.~(\ref{eq:intensity}) and (\ref{eq:O_Tr}) 
[see Eq.~(\ref{eq:Q-tilde})] is to make use of the following: 
if one denotes  by $E_j$, $j=1\dots n_d$, $n_d\le N$ the distinct  eigenvalues of a $N\times N$ 
hermitian matrix $H$, then the projector onto the $\eta$th eigenstate is given by the expression
\begin{equation}
Q^{\eta} = \frac{\prod\limits_{\eta\neq j}\left(H-E_j\hat{I}\right)}
{\prod\limits_{\eta\neq j}\left(E_{\eta}-E_j\right)}\;, \label{eq:proj}
\end{equation}
which does not necessitates the calculation of the eigenvectors. In the mathematical literature
the projectors $Q^{\eta}$ are known as Frobenius covariants\cite{frob-cov}. In terms of the projectors
$Q^{j}$ and eigenvalues $E_j$ the matrix $H$ is given by $H=\sum_{j=1}^{n_d}E_j Q^{j}$.

\bibliographystyle{prsty}

\begin{thebibliography}{10}
\bibitem{intensity-pattern}
Fig.1(d) of Ref.~\onlinecite{falko} shows for $\lambda_R=0$ 
the intensity pattern at a fixed energy covering several Brilloiun zones. 
A constant energy cross-section of the calculations 
shown in Fig.~\ref{fig:spinpol}(a) 
 at energies away from the  Dirac point 
(so that for this energy $k>\lambda_R/{\hbar v_F}$)  
would give a very similar result to the one in 
Fig.1(d) of Ref.~\onlinecite{falko}. Note however, that
because of the different choice of the coordinate
system, our $k_x$ ($k_y$) corresponds to $k_y$ ($k_x$) 
in  Ref.~\onlinecite{falko}.
\bibitem{lorentzian}
In Ref.~\cite{falko} a slightly different broadening 
function was considered: $\frac{1}{\pi}\frac{\Gamma}{\epsilon^2+\Gamma^2}$. 
We found that our choice is more convenient for the spin polarization 
calculations, the full width at half maximum of the functions, 
which is of  importance here, is  the same for both choices.
\end{thebibliography}

\end{document}